\documentclass[12pt]{article}
\usepackage{graphicx}
\newcommand{\be}{\begin{equation}}
\newcommand{\ee}{\end{equation}}
\newcommand{\ba}{\begin{eqnarray}}
\newcommand{\ea}{\end{eqnarray}}

\newcommand{\s}{\sqrt{2}}

\begin{document}

\title{{\bf Photon Boomerang in a \\ Nearly Extreme Kerr Metric}}

\author{
Don N. Page
\thanks{Internet address:
profdonpage@gmail.com}
\\
Department of Physics\\
4-183 CCIS\\
University of Alberta\\
Edmonton, Alberta T6G 2E1\\
Canada
}

\date{2021 June 24}

\maketitle
\large
\begin{abstract}
\baselineskip 20 pt

The Kerr rotating black hole metric has unstable photon orbits that orbit around the hole at fixed values of the Boyer-Lindquist coordinate $r$ that depend on the axial angular momentum of the orbit, as well as on the parameters of the hole.  For zero orbital axial angular momentum, these orbits cross the rotational axes at a fixed value of $r$ that depends on the mass $M$ and angular momentum $J$ of the black hole.  Nonzero angular momentum of the hole causes the photon orbit to rotate so that its direction when crossing the north polar axis changes from one crossing to the next by an angle I shall call $\Delta\phi$, which depends on the black hole dimensionless rotation parameter $a/M = cJ/(GM^2)$ by an equation involving a complete elliptic integral of the first kind.  When the black hole has $a/M \approx 0.994\,341\,179\,923\,26$, which is nearly maximally rotating, a photon sent out in a constant-$r$ direction from the north polar axis at $r \approx 2.423\,776\,210\,035\,73\, GM/c^2$ returns to the north polar axis in precisely the opposite direction (in a frame nonrotating with respect to the distant stars), a photon boomerang.

\end{abstract}

\normalsize

\baselineskip 22 pt

\newpage

\section{Introduction}

When a stationary observer sends a photon sufficiently near a black hole, it can orbit around the hole and come back to the observer.  However, after going around a nonrotating hole a certain number of times, the photon comes back to the observer from the opposite side of the hole from the direction at which it was sent by the observer, so it does not come back to the observer at exactly the opposite direction from that at which it was originally sent, in this or in any other static metric.  Thus it is not a photon boomerang, which I shall define as a photon orbit that returns to the stationary observer that sends it in precisely the opposite direction (in a reference frame nonrotating with respect to the distant stars in a stationary situation).

Here I shall show that a photon boomerang can occur for a rotating black hole with the Kerr metric (the unique stationary asymptotically flat vacuum metric in general relativity, with the Schwarzschild and Minkowski matrics being special cases), which has both mass $M$ and angular momentum $J$ with the dimensionless rotation parameter $a/M = cJ/(GM^2)$ having the allowed range from 0 to 1.  For the special value $a/M \approx 0.994\,341\,179\,923\,26$, there can be a photon boomerang that leaves the north polar axis ($\theta = 0$) at Boyer-Lindquist coordinate $r \approx 2.423\,776\,210\,035\,73\, GM/c^2$ and travels along a constant-$r$ orbit to return to the north polar axis in precisely the opposite direction to that at which it was emitted.  The dragging of inertial frames rotates the photon propagation direction by an angle $\Delta\phi = \pi$ during one orbit that returns to the same location on the north polar axis.

For an extreme Kerr black hole (maximally rotating with $a/M = 1$), the angle of rotation is twice the complete elliptical integral $K(m)$ with dimensionless argument $m \!=\! k^2 \!=\! \sin{\alpha} \!=\! [(\sqrt{2}-1)/2]^2$, giving $\Delta\phi \!=\! 2K([(\sqrt{2}-1)/2]^2) \!\approx\! 1.010\,989\,999\,412\,3\,\pi$.  When $a/M$ is reduced below its extremal value of 1, the prefactor, say $P$, of $K(m)$ drops below 2 (and the elliptic integral argument $m$ also decreases, though since it is small even for an extremal Kerr hole, this has less effect), so there is a value of $a/M$ slightly below 1 at which $\Delta\phi = P K(m) = \pi$, giving a photon boomerang.\\

Now I shall develop the equations giving the relations between $a/M$ and the elliptic integral parameter $m$ and prefactor $P$ and then solve for the value of $a/M$ giving $\Delta\phi = P K(m) = \pi$.  Along the way, I shall develop a more compact than usual infinite product for the complete elliptic integral of the first kind, $K(m)$, and, truncating it to the product of the first two factors, giving a rather short explicit closed-form algebraic expression for $K(m)$ that has a relative error always less than $4.04\times 10^{-41}$ for the possible Kerr black hole values for the elliptic integral argument
$m \leq [(\sqrt{2}-1)/2]^2 \approx 0.0429$, or for the much larger range of $m = \sin{\alpha}$ up to a modular angle of $\alpha = 89^\circ\, 59'\, 59.5''$ (e.g., for the amplitude of a simple pendulum up to $\theta_0 = 2\alpha = 179^\circ\,59'\, 59''$, whose period is $4\sqrt{\ell/g}K(m = \sin^2{\alpha} = (1-\cos{\theta_0})/2)$, the maximum error (at the upper end of this range for $m = \sin^2{\alpha}$) is about 1.591\%.

All of the numerical calculations of this paper were simple enough to be done on a pocket calculator without using any programmable features.  The ancient HP 48SX calculator I used gave 12-digit precision, and from the fact that $m/4 \sim 0.01$, sometimes I can get 2 extra digits, so usually I shall give the numerical results preceded by the $\approx$ sign to 12-14 digits, without checking whether the last digits are correct, though I suspect that in many cases they are, given the agreement I got in doing several of the calculations in two or more different ways.

\section{Constant-$r$ photon orbits in the Kerr metric}

Using units in which $G = c = 1$, the Kerr metric \cite{Kerr}, with mass $M$ and angular momentum $J = Ma \leq M^2$, in Boyer-Lindquist coordinates \cite{BL} is
\be
ds^2 = -dt^2 + (2Mr/\Sigma)(dt + a \sin^2{\theta}d\phi)^2 + (r^2+a^2)\sin^2{\theta}d\phi^2
+\Sigma(dr^2/\Delta + d\theta^2),
\ee
where
\be
\Sigma \equiv r^2 + a^2 \cos^2{\theta},\ \ \Delta \equiv r^2 - 2Mr + a^2.
\ee

Because the metric component functions are independent of both $t$ and $\phi$, the Kerr metric is both stationary and axisymmetric, so freely falling particles (moving along geodesics) have conserved energy $E = - p_t$ and angular momentum $L_z = p_\phi$, as well as the scalar product of the 4-momentum with itself, which is constant (minus the square of the rest mass of the particle) for any geodesic in any metric.  Carter \cite{Carter} showed that in the Kerr metric, geodesics have a fourth constant of motion, $\mathcal{K}$, making the geodesic equations of motion completely integrable.  However, since Carter used a slightly different coordinate system from the Boyer-Lindquist one, I shall take the detailed equations of motion from Misner, Thorne, and Wheeler (MTW) \cite{MTW}, which uses the Boyer-Lindquist coordinates that I am using, though note that I am following Boyer and Lindquist in using $\Sigma$ for $r^2 + a^2 \cos^2{\theta}$ instead of the $\rho^2$ that MTW uses.

For a photon crossing the polar axes, the axial angular momentum is zero, $L_z = p_\phi = 0$, and without generality for the geodesic trajectory of the photon, I shall take its energy to be 1, $E = - p_t = 1$.  Then the relevant equations for the spatial motion (not considering the time) from MTW Eqs. (33.32)-(33.33) on pages 899-900, with $\lambda$ being an affine parameter along the geodesic trajectory, are
\ba
\Sigma dr/d\lambda &=& \sqrt{R} \equiv \sqrt{(r^2+a^2)^2 - \mathcal{K}\Delta}, \label{dr}\\
\Sigma d\theta/d\lambda &=& \sqrt{\Theta} \equiv \sqrt{\mathcal{K} - a^2\sin^2{\theta}}, \label{dtheta}\\
\Sigma d\phi/d\lambda &=& 2Mar/\Delta. \label{dphi}
\ea

For the photon orbit to have constant $r$ in Eq.\ (\ref{dr}), it must be at a double root of $R(r) = (r^2+a^2)^2 - \mathcal{K}(r^2 - 2Mr + a^2)$, where $R = dR/dr = 0$.  This gives a unique value of $\mathcal{K}$ and of $r$ for a constant-$r$ photon orbit that crosses the polar axes ($\theta = 0$ and $\theta = \pi$).  The solution seems to be algebraically simplest if we let $r/M$ or some linear function of it be the independent variable and solve for $a/M$ and $\mathcal{K}/M^2$, both of which are dimensionless with my choice of $E = 1$.  In particular, for the later equations it seemed best to let the independent variable be
\be
y \equiv \frac{r-M}{M},
\ee
so that then
\ba
\frac{r}{M} &=& 1+y, \\
\frac{a^2}{M^2} &=& \frac{(2-y)(1+y)^2}{2+y}, \\
\delta \equiv 1 - \frac{a^2}{M^2} &=& \frac{y(y^2-2)}{2+y}, \label{cubic} \\
\frac{r^2+a^2}{M^2} &=& \frac{4(1+y)^2}{2+y}, \\
\frac{\Delta}{M^2} &=& \frac{2y(1+y)}{2+y}, \\
\frac{\mathcal{K}}{M^2} &=& \frac{8(1+y)^3}{y(2+y)}, \\
P \equiv \frac{8Mar}{\Delta\sqrt{\mathcal{K}}} &=& (2+y)\sqrt{\frac{2(2-y)}{y(1+y)}}, \label{P}\\
m \equiv \frac{a^2}{\mathcal{K}} &=& \frac{y(2-y)}{8(1+y)}, \label{m}\\
1-m &=& \frac{(2+y)(4+y)}{8(1+y)}.
\ea

If one first starts with $a/M$ or $\delta \equiv 1-(a/M)^2$, then one can solve the cubic equation $y^3-(2+\delta)y -2\delta = 0$ from Eq.\ (\ref{cubic}) to get $y = b\cos{\varphi}$ with $b = 2\sqrt{(2+\delta)/3}$ and $\cos{3\varphi} = \delta[3/(2+\delta)]^{3/2}$, with $3\varphi$ in the first quadrant, or
\ba
y \equiv \frac{r-M}{M} &=& 2\sqrt{\frac{2+\delta}{3}}\cos{\left\{\frac{1}{3}\cos^{-1}{\left[\delta\left(\frac{3}{2+\delta}\right)^{3/2}\right]}\right\}} \nonumber \\
&=& 2\sqrt{\frac{3M^2-a^2}{3M^2}}
\cos{\left\{\frac{1}{3}\cos^{-1}{\left[\left(1-\frac{a^2}{M^2}\right)\left(\frac{3M^2}{3M^2-a^2}\right)^{3/2}\right]}\right\}} \label{ysolution}.
\ea
For $a=0$ or $\delta = 1$, $3\varphi = 0$ and $y=2$, so $r = M(1+y) = 3M$, the circular photon orbit in the Schwarzschild metric.  For $a=M$ or $\delta = 0$, $3\varphi = \pi/2$ and $y = \sqrt{2}$, or $r = M(1+y) = (\sqrt{2}+1)M$ for the extreme Kerr black hole.  Thus $y$ decreases from $2$ to $\sqrt{2}$ as $a/M$ increases from 0 to 1.

Now I shall integrate the ratio of Eqs.\ (\ref{dphi}) and (\ref{dtheta}) to show that the change in $\phi$ as the photon goes from the north polar axis ($\theta = 0$) to the south polar axis ($\theta = \pi$) and then back to the north polar axis is $\Delta\phi = P K(m)$, where $P$ and the argument $m = k^2 = \sin^2{\alpha}$ (where $k = \sqrt{m} = \sin{\alpha}$ is the modulus and $\alpha$ is the modular angle) of the complete elliptic integral of the first kind, $K(m)$, are given in terms of $y$ by Eqs.\ (\ref{P}) and (\ref{m}) respectively.  The ratio of Eqs.\ (\ref{dphi}) and (\ref{dtheta}) is
\be
\frac{d\phi}{d\theta} 
= \frac{2Mar/\Delta}{\sqrt{\mathcal{K} - a^2\sin^2{\theta}}} 
= \frac{P(y)}{4\sqrt{1-m\sin^2{\theta}}}.
\ee
Since
\be
K(m) \equiv \int_0^{\pi/2}\frac{d\theta}{\sqrt{1-m\sin^2{\theta}}},
\ee
\be
\Delta\phi = 2\int_0^\pi \frac{d\phi}{d\theta} d\theta = P K(m) \label{Deltaphi}.
\ee
I am not counting the two sudden changes of $\phi$ by $\pi$ radians when the photon crosses the north and south polar axes.

I shall first say what happens for small $a/M$.  Then
\ba
y \equiv \frac{r-M}{M} &=& 2-\frac{4}{9}\left(\frac{a}{M}\right)^2-\frac{20}{243}\left(\frac{a}{M}\right)^4
-\frac{196}{6\,561}\left(\frac{a}{M}\right)^6 +O\left(\frac{a^8}{M^8}\right), \\
\frac{r}{M} &=& 3-\frac{4}{9}\left(\frac{a}{M}\right)^2-\frac{20}{243}\left(\frac{a}{M}\right)^4
-\frac{196}{6\,561}\left(\frac{a}{M}\right)^6 +O\left(\frac{a^8}{M^8}\right), \\
\frac{r^2+a^2}{M^2} &=& 9-\frac{5}{3}\left(\frac{a}{M}\right)^2-\frac{8}{27}\left(\frac{a}{M}\right)^4
 +O\left(\frac{a^6}{M^6}\right), \\
\frac{\Delta}{M^2} &=& 3-\frac{7}{9}\left(\frac{a}{M}\right)^2-\frac{32}{243}\left(\frac{a}{M}\right)^4 +O\left(\frac{a^6}{M^6}\right), \\
\frac{\mathcal{K}}{M^2} &=& 27-3\left(\frac{a}{M}\right)^2-\frac{4}{9}\left(\frac{a}{M}\right)^4 
+O\left(\frac{a^6}{M^6}\right), \\
m \equiv \frac{a^2}{\mathcal{K}} &=& \frac{1}{27}\left(\frac{a}{M}\right)^2
+\frac{1}{243}\left(\frac{a}{M}\right)^4
+\frac{7}{6\,561}\left(\frac{a}{M}\right)^6 +O\left(\frac{a^8}{M^8}\right), \\
P \equiv \frac{8Mar}{\Delta\sqrt{\mathcal{K}}} &=& \frac{8}{\sqrt{27}}\left(\frac{a}{M}\right)
\left[27-3\left(\frac{a}{M}\right)^2-\frac{4}{9}\left(\frac{a}{M}\right)^4 
+O\left(\frac{a^6}{M^6}\right)\right], \\
K(m) &=& \frac{\pi}{2}\left[1+\frac{1}{108}\left(\frac{a}{M}\right)^2-\frac{19}{15\,552}\left(\frac{a}{M}\right)^4 
+O\left(\frac{a^6}{M^6}\right)\right], \\
\Delta\phi = P K(m) \!\!&=&\!\! \frac{4\pi}{\sqrt{27}}\!\left(\frac{a}{M}\right)\!\left[1\!+\!\frac{19}{108}\left(\frac{a}{M}\right)^2\!+\!\frac{1\,031}{15\,552}\left(\frac{a}{M}\right)^4 
\!+O\left(\frac{a^6}{M^6}\right)\right].
\ea

On the other hand, near the extreme Kerr metric, one can expand in powers of $\delta \equiv 1 - (a/M)^2$ to get
\newpage
\ba
y \equiv \frac{r-M}{M} &=& \s+\frac{2+\s}{4}\,\delta-\frac{7\s+8}{32}\,\delta^2
+\frac{48+31\s}{128}\,\delta^3 +O\left(\delta^4\right)\!, \\
\frac{r}{M} &=& \s+1\!+\!\frac{2+\s}{4}\,\delta\!-\!\frac{7\s+8}{32}\,\delta^2
\!+\!\frac{48+31\s}{128}\,\delta^3 \!+\!O\left(\delta^4\right)\!\!, \\
\frac{r^2+a^2}{M^2} &=& 4+2\s+\frac{3\s+2}{2}\,\delta-\frac{16+11\s}{16}\,\delta^2
 +O\left(\delta^3\right), \\
\frac{\Delta}{M^2} &=& 2+\s\,\delta-\frac{2+\s}{4}\,\delta^2 +O\left(\delta^3\right), \\
\frac{\mathcal{K}}{M^2} &=& (12+8\s)+2(\s+1)\,\delta-\frac{3\s+4}{4}\,\delta
+O\left(\delta^3\right), \\
m \equiv \frac{a^2}{\mathcal{K}} &=& \frac{3-2\s}{4}-\frac{\s-1}{8}\,\delta
+\frac{8-5\s}{64}\,\delta +O\left(\delta^3\right), \\
P \equiv \frac{8Mar}{\Delta\sqrt{\mathcal{K}}} &=& 2-\frac{2\s+1}{2}\,\delta
+\frac{15+4\s}{16}\,\delta +O\left(\delta^3\right), \\
K(m) &=& K_m +\left(\frac{\s+1}{4}K_m-\frac{3+\s}{7}E_m\right)\delta \nonumber \\
 &+& \left(\frac{15+19\s}{224}K_m-\frac{52+15\s}{392}E_m\right)\delta^2 +O\left(\delta^3\right),\label{Km} \\
\Delta\phi = P K(m) &=& 2K_m - \left(\frac{\s}{2}K_m+\frac{6+2\s}{7}E_m\right)\delta \nonumber \\
 &+& \left(\frac{5(10+\s)}{112}K_m+\frac{46+83\s)}{196}E_m\right)\delta^2 
+O\left(\delta^3\right) \label{Dphi}.
\ea

Here $K_m$ and $E_m$ are the values of the complete elliptic integrals of the first and second kind, respectively, evaluated at the parameter $m = a^2/\mathcal{K}$ value at $a=M$, namely $m_m = [(\s-1)/2]^2 = (3-2\s)/4 \approx 0.042\,893\,218\,813\,4$:
\be
K_m \equiv K(m_m) = K\left(\frac{3-2\s}{4}\right)\ 
\mathrm{and}\ E_m \equiv E(m_m) = E\left(\frac{3-2\s}{4}\right).
\ee
The Taylor expansion in Eq.\ (\ref{Km}) uses the formulas for the derivatives of the complete elliptic integrals with respect to their parameter $m = k^2 = \sin^2{\alpha}$:
\ba
\frac{dK(m)}{dm} &=& \frac{E(m)}{2m(1-m)}-\frac{K(m)}{2m}, \\
\frac{dE(m)}{dm} &=& \frac{E(m)-K(m)}{2m}, \\
\frac{d^2K(m)}{dm^2} &=& \frac{2-3m}{4m^2(1-m)}K(m) - \frac{1-2m}{2m^2(1-m)^2}E(m).
\ea

\section{Short closed-form approximate formulas for the elliptic integrals}

I have shown that the angular rotation (change in direction in a frame that is nonrotating as seen from infinity) of a photon in a polar orbit with constant-$r$ during one orbit is given by Eq.\ (\ref{Deltaphi}) as $\Delta\phi = P K(m)$, where Eqs.\ (\ref{P}) and (\ref{m}) give the prefactor $P$ and the parameter $m$ of the complete elliptical integral $K(m)$ in terms of $y=(r-M)/M$ given in terms of $a/M$ or $\delta\equiv 1-(a/M)^2$ by Eq.\ (\ref{ysolution}) that is the solution to the cubic equation Eq.\ (\ref{cubic}).  To get fairly compact explicit closed-form approximations in terms of elementary functions for $\Delta\phi$ and to use them to solve for the values of $y$ and of $a/M$ that give the photon boomerang, $\Delta\phi = \pi$, I shall first show how to get compact closed-form approximations for the complete elliptic integral $K(m)$ (and also ones for $E(m)$) when $m$ is not too large, as is the case for the constant-$r$ photon polar orbits in the Kerr metric, for which $m \leq m_m = [(\s-1)/2]^2 = (3-2\s)/4 \approx 0.042\,893\,218\,813\,4$.

One of the fastest converging sequences for getting the complete elliptic integrals $K(m)$ and $E(m)$ is to use the arithmetic-geometric mean \cite{AS}, using these initial values and recursion: 
\ba
a_0 &=& 1,\\
b_0 &=& \cos{\alpha} = \sqrt{1-m},\\
c_0 &=& \sin{\alpha} = \sqrt{m}\,;\\
a_{n+1} &=& \frac{1}{2}(a_n+b_n),\\
b_{n+1} &=& \sqrt{a_n b_n},\\
c_{n+1} &=& \frac{1}{2}(a_n-b_n).
\ea
This gives a sequence
\be
K_n(m) = \frac{\pi}{2a_n}
\ee
of better and better approximations for $K(m)$, which one can terminate when $a_n$ and $b_n$ agree for the desired number of significant digits of $K(n)$.  For evaluating $K(m)$, the $c_n$'s are unnecessary, but they are needed for evaluating $E(m)$ as the limit of the sequence
\be
E_n(m) = K_n(m)[1-(1/2)c_0^2-c_1^2-2c_2^2-2^2c_3^2-\cdots -2^{n-2}c_{n-1}^2]. \label{E/K}
\ee

It is convenient to do two iterations at once, by defining
\ba 
\alpha_n &\equiv& \sqrt{a_n}\ \mathrm{(not\ to\ be\ confused\ with\ the\ modular\ angle}\ \alpha),\\
\beta_n &\equiv& \sqrt{b_n},\\
\gamma_n &\equiv& \sqrt{c_n},
\ea
with initial values and compressed recursion relation
\ba
\alpha_0 &=& 1,\\
\beta_0 &=& \sqrt{\cos{\alpha}} = (1-m)^{1/4},\\
\gamma_0 &=& \sqrt{\sin{\alpha}} = m^{1/4}; \\
\alpha_{2n+2} &=& \frac{1}{2}(\alpha_{2n}+\beta_{2n}),\\
\beta_{2n+2} &=& [(1/2)\alpha_{2n}\beta_{2n}(\alpha_{2n}^2+\beta_{2n}^2)]^{1/4},\\
\gamma_{2n+2} &=& \frac{1}{2}(\alpha_{2n}-\beta_{2n}).
\ea
If one defines
\be
\delta_{2n} \equiv \frac{\gamma_{2n}}{\alpha_{2n}} \equiv \sqrt{\frac{c_{2n}}{a_{2n}}},
\ee
(not to be confused with $\delta \equiv 1 - (a/M)^2$), then one gets this initial condition and recursion relation for a single sequence of $\delta_n$'s with just even values of $n$:
\ba
\delta_0 &=& m^{1/4} = \sqrt{\sin{\alpha}}, \label{delta0} \\
\delta_{2n+2} &=& \frac{1-(1-\delta_{2n}^4)^{1/4}}{1+(1-\delta_{2n}^4)^{1/4}}
= \frac{\delta_{2n}^4}{[1+(1-\delta_{2n}^4)^{1/4}]^2[1+\sqrt{1-\delta_{2n}^4}]}. \label{delta2n}
\ea
This leads the following sequence of rapidly improving approximations for $K(m)$:
\be
K_{2n}(m) = \frac{\pi}{2\alpha_{2n}^2} = \frac{\pi}{2}\prod_{i=1}^{n}(1+\delta_{2i})^2. \label{K2n}
\ee

Alternatively, one can get a two-term recursion relation for the $\alpha_{2n}$'s and for the $K_{2n}(m)$'s:
\ba
\alpha_{2n+4} &=& \frac{1}{2}\alpha_{2n+2}+\frac{1}{2}\left[\alpha_{2n+2}^4-(\alpha_{2n}-\alpha_{2n+2})^2\right]^{1/4},\\
K_{2n+4} &=& K_{2n+2}\left\{\frac{1}{2}+\frac{1}{2}\left[1-\left(\sqrt{\frac{K_{2n+2}}{K_{2n}}}-1\right)^4\right]^{1/4}\right\}^{\!\!-2}.
\ea

To relocate some factors of $1/2$ to give slightly shorter expressions, one can define
\ba
A_{2n} &\equiv& 2^n\alpha_{2n} \equiv 2^n\sqrt{a_{2n}},\\ 
B_{2n} &\equiv& 2^n\beta_{2n} \equiv 2^n\sqrt{b_{2n}},\\  
C_{2n} &\equiv& 2^n\gamma_{2n} \equiv 2^n\sqrt{c_{2n}}.
\ea
These give
\be
K_{2n}(m) = \frac{\pi}{2}\left(\frac{2^n}{A_{2n}}\right)^2
\ee
and have the order-1 recursion relations
\ba
A_{2n+2} &=& A_{2n}+B_{2n},\\
C_{2n+2} &=& A_{2n}-B_{2n},\\
B_{2n+2} &=& (A_{2n+2}^4 - C_{2n+2}^4)^{1/4} = [8A_{2n}B_{2n}(A_{2n}^2+B_{2n}^2)]^{1/4}.
\ea
The order-2 recursion relation for $A_{2n}$ is
\be
A_{2n+4} = A_{2n+2} + [A_{2n+2}^4 - (A_{2n}-A_{2n+2})^4]^{1/4}.
\ee

In particular, if for brevity one defines
\be
\beta \equiv \beta_0 = (1-m)^{1/4} = \sqrt{\cos{\alpha}}, \label{beta}
\ee
then
\newpage
\ba
\delta_0 \!\!\!\! &=&\!\!\!\! m^{1/4} = \sqrt{\sin{\alpha}} = (1-\beta^4)^{1/4},\\
\delta_2 \!\!\!\! &=&\!\!\!\! \frac{1-\beta}{1+\beta} = \frac{m}{(1+\beta)^2(1+\beta^2)},\\ 
\delta_4 \!\!\!\! &=&\!\!\!\! \frac{1+\beta-[8\beta(1+\beta^2)]^{1/4}}{1+\beta+[8\beta(1+\beta^2)]^{1/4}},\\
\delta_6 \!\!\!\! &=&\!\!\!\!\! \frac{1\!\!+\!\!\beta\!\!+\!\![8\beta(1\!+\!\beta^2)]^{1/4}\!-\!\{8(1\!+\!\beta)[8\beta(1\!+\!\beta^2)]^{1/4}[(1\!+\!\beta)^2\!+\!\sqrt{8\beta(1\!+\!\beta^2)}]\}^{1/4}}
{1\!\!+\!\!\beta\!\!+\!\![8\beta(1\!+\!\beta^2)]^{1/4}\!+\!\{8(1\!+\!\beta)[8\beta(1\!+\!\beta^2)]^{1/4}
[(1\!+\!\beta)^2\!+\!\sqrt{8\beta(1\!+\!\beta^2)}]\}^{1/4}},\\
A_0  \!\!\!\! &=&\!\!\!\!  1,\\
A_2  \!\!\!\! &=&\!\!\!\!  1+\beta,\\
A_4  \!\!\!\! &=&\!\!\!\!  1+\beta+[8\beta(1+\beta^2)]^{1/4},\\
A_6  \!\!\!\! &=&\!\!\!\!  1\!+\!\beta\!+\![8\beta(1\!+\!\beta^2)]^{1/4}\!+\!\{8(1\!+\!\beta)[8\beta(1\!+\!\beta^2)]^{1/4}[(1\!+\!\beta)^2\!+\!\sqrt{8\beta(1\!+\!\beta^2)}]\}^{\!1/4}\!\!,\\
C_0 \!\!\!\! &=&\!\!\!\! (1-\beta^4)^{1/4} = m^{1/4} = \sqrt{\sin{\alpha}},\\
C_2 \!\!\!\! &=&\!\!\!\! 1-\beta,\\
C_4 \!\!\!\! &=&\!\!\!\! 1+\beta-[8\beta(1+\beta^2)]^{1/4},\\
C_6 \!\!\!\! &=&\!\!\!\! 1\!+\!\beta\!+\![8\beta(1\!+\!\beta^2)]^{1/4}\!-\!\{8(1\!+\!\beta)[8\beta(1\!+\!\beta^2)]^{1/4}[(1\!+\!\beta)^2\!+\!\sqrt{8\beta(1\!+\!\beta^2)}]\}^{1/4}\!\!,\\ 
B_0 \!\!\!\! &=&\!\!\!\! \beta,\\
B_2 \!\!\!\! &=&\!\!\!\! [(1+\beta)^4-(1-\beta)^4]^{1/4} = [8\beta(1+\beta^2)]^{1/4},\\
B_4 \!\!\!\! &=&\!\!\!\! \{8(1\!+\!\beta)[8\beta(1\!+\!\beta^2)]^{1/4}[(1+\beta)^2+\sqrt{8\beta(1+\beta^2)}]\}^{1/4},\\
B_6 \!\!\!\! &=&\!\!\!\! (A_6^4-C_6^4)^{1/4} = [8A_4B_4(A_4^2+B_4^2)]^{1/4}.
\ea
This then leads to closed forms for approximations to $K(m)$ for $m$ not too large:
\ba
K_0(m)\!\!\!\! &=&\!\!\!\! \frac{\pi}{2}, \label{K0} \\
K_2(m)\!\!\!\! &=&\!\!\!\! \frac{\pi}{2}\!\left(\!\frac{2}{1\!+\!\beta}\!\right)^2
 \!= \frac{\pi}{2}\!\left(\!\frac{2}{1\!+\!(1\!-\!m)^{1/4}}\!\right)^{\!\!2}
 \!= \frac{\pi}{2}\!\left(\!\frac{2}{1\!+\!\sqrt{\cos{\alpha}}}\!\right)^{\!\!2}
 \!= \frac{\pi}{2}\!\left[1\!+\!\frac{(3\!+\!\beta)m}{(1\!+\!\beta)^3(1\!+\!\beta^2)}\!\right]\!\!, \label{K2}\\
K_4(m)\!\!\!\! &=&\!\!\!\! \frac{\pi}{2}\left(\frac{4}{1+\beta+[8\beta(1+\beta^2)]^{1/4}}\right)^{\!\!2}
\! = \frac{\pi}{2}\!\left(\frac{4}{1\!+\!\sqrt{\cos{\alpha}}\!+\!2(\cos{\alpha})^{1/8}\sqrt{\cos{(\alpha/2)}}}\right)^{\!\!2}\!\!, \label{K4}\\
K_6(m)\!\!\!\!\! &=&\!\!\!\!\! \frac{\pi}{2}\!\left(\!\frac{8}{1\!+\!\beta\!+\![8\beta(1\!+\!\beta^2)]^{1/4}\!+\!\{8(1\!+\!\beta)[8\beta(1\!+\!\beta^2)]^{1/4}
[(1\!+\!\beta)^2\!+\!\sqrt{8\beta(1\!+\!\beta^2)}]\}^{1/4}}\!\right)^{\!\!2}. \nonumber
\ea

For getting approximations for the complete elliptical integral of the second kind, $E(m)$, we can use Eq.\ (\ref{E/K}).  Using
\ba
c_0 &=& \sqrt{m} = \sin{\alpha} = \sqrt{1-\beta^4},\\
c_1 &=& \frac{1}{2}(1-\beta^2),\\
c_2 &=& \frac{1}{4}(1-\beta)^2,\\
c_3 &=& \frac{1}{8}\left[(1+\beta)^2 - \sqrt{8\beta(1+\beta^2)}\right],\\
c_4 &=& \frac{1}{16}\left\{1+\beta - [8\beta(1+\beta^2)]^{1/4}\right\}^{\!\!2},\\
c_5 &=& \frac{1}{32}\left\{\sqrt{(1+\beta)^2+\sqrt{8\beta(1+\beta^2)}}-[32\beta(1+\beta^2)]^{1/4}\right\}^{\!\!2},
\ea
one gets
\ba
E_0 \!\!\!\! &=&\!\!\!\! K_0 = \frac{\pi}{2},\\
E_2 \!\!\!\! &=&\!\!\!\! K_2\left(1-\frac{1}{2}c_0^2-c_1^2\right) = \frac{\pi}{2}\left(\frac{1+\beta^2}{1+\beta}\right)^{\!\!2}
= \frac{\pi}{2}\left(\frac{1+\cos{\alpha}}{1+\sqrt{\cos{\alpha}}}\right)^{\!\!2},\\
E_4 \!\!\!\! &=&\!\!\!\! K_4\!\left(\!1\!-\!\frac{1}{2}c_0^2\!-\!c_1^2\!-\!2c_2^2\!-\!4c_3^2\!\right) 
 \!=\! \frac{\pi}{2}\frac{(1\!+\!\beta)^2\left[1\!-\!6\beta\!+\!\beta^2\!+\!2\sqrt{8\beta(1\!+\!\beta^2)}\right]}{\left\{1\!+\!\beta\!+\![8\beta(1\!+\!\beta^2)]^{1/4}\right\}^2}\!, \label{E4}\\
E_6 \!\!\!\! &=&\!\!\!\! K_6\left(1-\frac{1}{2}c_0^2-c_1^2-2c_2^2-4c_3^2-8c_4^2-16c_5^2\right).
\ea

Now let us estimate the relative errors of the estimates $K_{2n}$ and $E_{2n}$, say
\ba
\frac{\Delta K_{2n}(m)}{K(m)} &\equiv& \frac{K_{2n}(m)-K(m)}{K(m)},\\
\frac{\Delta E_{2n}(m)}{E(m)} &\equiv& \frac{E_{2n}(m)-E(m)}{E(m)}.
\ea
From Eqs.\ (\ref{K2n}) and (\ref{delta2n}), we see that the relative error of $K_{2n}(m)$ is
\be
\frac{\Delta K_{2n}(m)}{K(m)} = - 2\delta_{2n+2} + O(\delta_{2n+2}^2) = -\frac{1}{4}\delta_{2n}^4 + O(\delta_{2n}^8),
\ee
so when $\delta_{2n}$ is small, $\delta_{2n+2}$ and $\Delta K_{2n}(m)/K(m)$ are very small.  For any $m < 1$, the iterations of $\delta_{2n}$ given by Eq.\ (\ref{delta2n}) eventually lead to $\delta_{2n} \ll 1$, and then the errors rapidly get smaller with $n$, decreasing quartically so that the next iteration after $\delta_{2n}$ gets small goes as the 4th power of the previous error.  If $m$ itself is small, as it is for the constant-$r$ photon orbits through the polar axes for the Kerr metric, then, starting with $\delta_0 = m^{1/4}$ as given by Eq.\ (\ref{delta0}), one can show inductively that the relative error of $K_{2n}(m)$ is approximately
\be
\frac{\Delta K_{2n}(m)}{K(m)} \sim -4\left(\frac{m}{16}\right)^{2^{2n}}
 = -2^{-2\left(2^{2n+1}-1\right)} m^{2^{2n}}.
\ee
In particular, one has, for $m \ll 1$,
\ba
\frac{\Delta K_{0}(m)}{K(m)} &\sim& -2^{-2}m = -\frac{1}{4}m,\\
\frac{\Delta K_{2}(m)}{K(m)} &\sim& -2^{-14}m^4 = -\frac{1}{16\,384} m^4,\\
\frac{\Delta K_{4}(m)}{K(m)} &\sim& -2^{-62}m^{16},\\
\frac{\Delta K_{6}(m)}{K(m)} &\sim& -2^{-254}m^{64}.
\ea
Similarly, for $E_{2n}$ given by Eq.\ (\ref{E/K}), $K_{2n}(m)$ has the relative error given above, and the quantity in the square brackets that is $E_{2n}(m)/K_{2n}(m)$ has, when it is close to unity, the relative error $\sim 2^{2n-1} c_{2n}^2 \sim 2^{2n-1}\delta_{2n}^4$.  This then leads to the following estimate for the relative error for these approximations for the complete elliptic integral of the second kind, $E(m)$, when its parameter $m$ is sufficiently small:
\be
\frac{\Delta E_{2n}(m)}{E(m)} \sim \left(2^{2n+1}-1\right)4\left(\frac{m}{16}\right)^{2^{2n}}
 = \left(2^{2n+1}-1\right)2^{-2\left(2^{2n+1}-1\right)} m^{2^{2n}}.
\ee
For this complete elliptical integral, one thus has, again for $m \ll 1$,
\ba
\frac{\Delta E_{0}(m)}{E(m)} &\sim& +1\cdot 2^{-2}m = +\frac{1}{4}m,\\
\frac{\Delta E_{2}(m)}{E(m)} &\sim& +7\cdot 2^{-14}m^4 = +\frac{7}{16\,384} m^4,\\
\frac{\Delta E_{4}(m)}{E(m)} &\sim& +31\cdot 2^{-62}m^{16},\\
\frac{\Delta E_{6}(m)}{E(m)} &\sim& +127\cdot 2^{-254}m^{64}.
\ea
Note that the integer prefactor is $-1/2$ times the exponent of the power of 2.

For the extreme Kerr black hole with $a=M$ and hence $m$ taking its maximum value, $m_m = [(\s-1)/2]^2 = 1/(12+\sqrt{128}) \approx 0.042\,893\,218\,813\,4$, we get
\ba
\frac{\Delta K_{0}(m)}{K(m)} &\approx& -(1.013\,729\,682\,0)\,2^{-2}m \approx -0.010\,875\,322\,67,\\
\frac{\Delta K_{2}(m)}{K(m)} &\approx& -(1.091\,246\,666\,2)\,2^{-14}m^4 \approx -2.254\,536\,764\,7\times 10^{-10},\\
\frac{\Delta K_{4}(m)}{K(m)} &\approx& -(1.418\,047\,692\,08)\,2^{-62}m^{16} \approx -4.036\,912\,295\times 10^{-41},\\
\frac{\Delta K_{6}(m)}{K(m)} &\approx& -(4.043\,554\,952)\,2^{-254}m^{64} \approx - 4.149\,705\,559\times 10^{-164}.
\ea
Thus, for this small but not extremely small value of $m$, the relative errors are a few percent more than the earlier approximations for $m \ll 1$ for $K_0$ and $K_2$ and are within an order of magnitude of unity for $K_4$ and $K_6$, though for the latter two the errors are so small that the corrections to the crude estimates earlier are not likely to be important.

Since $K_2(m)$ has somewhat more error than we might want for, say, 12-digit accuracy, whereas $K_4(m)$ (if evaluated precisely) is much more precise than might be needed for $m \leq m_m$, it may be useful to use the following rational (in $\beta = (1-m)^{1/4} = \sqrt{\cos{\alpha}}$) approximations for $K_4(m)$:
\ba
\bar{K}_4(m) &=& \frac{\pi}{2}\left[\frac{4(1+\beta)^4+(1-\beta)^4}{(1+\beta)^6}\right] \label{Kbar1}\\
&=& \frac{\pi}{2}\left[1+\frac{(3+\beta)m}{(1+\beta)^3(1+\beta^2)}+\frac{m^4}{(1+\beta)^{10}(1+\beta^2)^4}\right] \label{Kbar2} \\
&=& \frac{\pi}{2}\left[1+\frac{(4+10\beta+25\beta^2+17\beta^3+7\beta^4+\beta^5)m}{(1+\beta)^7(1+\beta^2)}
\right] \label{Kbar3} \\
&\approx& \frac{\pi}{2}\left[1-\frac{(13+25\beta+23\beta^2+3\beta^3)m}{64(1+\beta)^3(1+\beta^2)}\right]^{-1}
\label{last}.
\ea
The expressions after the second, third, and fourth equal signs are useful for getting $(2/\pi)\bar{K}_4(m)-1$ or $\pi/(2\bar{K}_4(m))-1$ to high accuracy when $m\ll 1$.  (Retaining just the first two terms inside the square brackets of the expression after the second equal sign gives $K_2(m)$.)  When $m^4\ll 1$, the relative error of all but the last of these approximations for $K(m)$ is
\be
\frac{\Delta \bar{K}_{4}(m)}{K(m)} \approx 
-9\cdot 2^{-30}\left(\frac{2}{1+\beta}\right)^{16}\left(\frac{2}{1+\beta^2}\right)^8 m^8, \label{relerr}
\ee
which for $m = m_m = [(\s-1)/2]^2$ is approximately $-(1.19081807574)9\cdot 2^{-30}m^8 \approx -1.143660606\times 10^{-19}$, which is certainly good enough for high-precision calculations of the photon boomerang in the nearly extreme Kerr metric.  For the last of these approximations, Eq (\ref{last}), the relative error is that given in Eq. (\ref{relerr}), with the $-9$ replaced by $-7$ (that is, 7/9 as large).

For an analogous approximation $\bar{E}_4(m)$ that is a rational function of my $\beta = (1-m)^{1/4} = \sqrt{\cos{\alpha}}$ and has error only $O(m^8)$, use Eq.\ (\ref{E4}) with $K_4$ replaced by $\bar{K}_4$ and with the $-4c_3^2$ term, which is $O(m^8)$, dropped:
\ba
\bar{E}_4(m) \!\!=\!\! \frac{\pi}{2}\left[\frac{4(1+\beta)^4+(1-\beta)^4}{(1+\beta)^6}\right]
\left[1-\frac{1}{8}(1-\beta)(7+3\beta+5\beta^2+\beta^3)\right] \\
\!\!=\!\!  \frac{\pi}{2}\!\left[1\!+\!\frac{(3\!+\!\beta)m}{(1\!+\!\beta)^3(1\!+\!\beta^2)}\!+\!\frac{m^4}{(1\!+\!\beta)^{10}(1\!+\!\beta^2)^4}\right]\!\!\!\left[1\!-\!\frac{(3\!+\!\beta^2)m}{4(1\!+\!\beta^2)}\!-\!\frac{m^4}{8(1\!+\!\beta)^4(1\!+\!\beta^2)^4}\right]\!\! \label{Ebar}.
\ea
When $m^4\ll 1$, the relative error of this approximation for $E(m)$ is
\be
\frac{\Delta \bar{E}_{4}(m)}{E(m)} \approx 
+7\cdot 2^{-30}\left(\frac{2}{1+\beta}\right)^{16}\left(\frac{2}{1+\beta^2}\right)^8 m^8,
\ee
which is of the same general magnitude (but opposite sign) of the relative error of $\bar{K}_4(m)$.

\section{Using the elliptic integral approximations for calculating the photon boomerang}

Now let us find the value of $y = (r-M)/M$, $\delta \equiv 1-(a/M)^2 = y(y^2-2)/(2+y)$, $r = M(1+y)$, and $a/M = \sqrt{1-\delta} = \sqrt{(2-y)(1+y)^2/(2+y)}$ at which the rotation of the direction of the photon during one orbit (return to the north polar axis) is $\Delta = P K[m] = (2+y)\sqrt{2(2-y)/[y(1+y)]}K[(1/8)y(2-y)/(1+y)] = \pi$.  I shall start with the Taylor-series expansion of Eq.\ (\ref{Dphi}) for $\delta \equiv 1-(a/M)^2 \ll 1$ but then use Newton's method to home in on the values of $y$ etc.\ that make $\Delta = \pi$.

Using the approximations of Eq.\ (\ref{Kbar1})-(\ref{Ebar}) for $m = m_m = (\s-1)^2/4 = 0.75-\sqrt{0.5} = 1/(12+\sqrt{128})$ (this last expression best for getting the most precision from a pocket calculator, giving $m_m = 0.042\,893\,218\,813\,4$ on my HP 48X that gives 12 significant digits, whereas subtracting by hand the value given by \cite{AS}, p.\ 2, for $2^{-1/2}$ from 0.75 gives $0.042\,893\,218\,813\,452$ to 15 digits after the decimal point, so the last digit given by the HP 48X is incorrectly rounded), results in
\ba
\frac{2}{\pi}K_m = \frac{2}{\pi}K(m_m) = K\left(\frac{3-2\s}{4}\right) &\approx& 1.100\,989\,999\,412\,3, \\
\frac{2}{\pi}E_m = \frac{2}{\pi}E(m_m) = E\left(\frac{3-2\s}{4}\right) &\approx& 0.989\,188\,874\,855\,5.
\ea
I can get roughly two more digits than the 12 significant digits given by my HP 48X calculator by calculating $\frac{2}{\pi}K_m - 1$ and $\frac{2}{\pi}E_m - 1$ (which are both of the order of $0.01$) and adding the 1 by hand, though again the last digit (or possibly the last two digits) might be incorrect because of rounding errors, as may other numerical values I give below, though I shall not keep repeating this warning.  It shall be left as an exercise for the reader to find which of the last digits are incorrect.

Inserting these values into Eq.\ (\ref{Dphi}) gives
\be
\frac{\Delta\phi}{\pi} = \frac{1}{2}P \frac{2}{\pi} K \approx
 1.010\,989\,999\,412\,3 - 0.981\,223\,363\,166\delta + 0.669\,860\,701\,304\delta^2 + O(\delta^3). \label{quad}
\ee
Ignoring the unwritten $O(\delta^3)$ terms and solving the resulting quadratic equation to get $\delta$ that would give $\Delta\phi/\pi = 1$ with this approximation results in
\be
\delta \approx 0.011\,287\,278\,576\,6.
\ee
Then from Eq.\ (\ref{ysolution}), one gets
\be
y \approx 1.423\,777\,606\,98.
\ee
This in turn by Eq.\ (\ref{P}) gives
\be
\frac{1}{2}P = (2+y)\sqrt{\frac{2-y}{2y(1+y)}} \approx 0.989\,277\,545\,710,
\ee
and by Eq.\ (\ref{m}) gives
\be
m = \sin^2{\alpha} = \frac{y(2-y)}{8(1+y)} \approx 0.042\,310\,634\,103\,8,
\ee
which leads by Eq.\ (\ref{beta}) to
\be
\beta = (1-m)^{1/4} \approx 0.989\,250\,244\,482.
\ee
Plugging these values of $m$ and $\beta$ into Eq.\ (\ref{Kbar3}) then gives
\be
\frac{2}{\pi}K(m) \approx 1.010\,837\,048\,972\,1,
\ee
and multiplying this by $P/2 \approx 0.989\,277\,545\,710$ yields
\be
\frac{\Delta\phi}{\pi} \approx 0.999\,998\,394\,918,
\ee
which is very close to 1, differing by only about $1.6\times 10^{-6}$, so already from the quadratic approximation (in $\delta = 1 - (a/M)^2$) of Eq.\ (\ref{Dphi}) or (\ref{quad}), we can get a reasonably good approximation for the values of $\delta$, $y$, etc. at which $\Delta\phi = \pi$.

We can get a much more precise value for $y$ at which $\Delta\phi = \pi$ by using Newton's method, say for finding the relevant root of
\be
L(z) \equiv \ln{\frac{\Delta\phi}{\pi}} = \ln{\frac{PK}{\pi}} = 0
\ee
by a sequence of iterations 
\be
z_{i+1} = z_i - L(z_i)/L'(z_i),
\ee
where 
\ba
L'(z) \!\!\!\! &\equiv&\!\!\!\! \frac{dL}{dz} = \frac{dP}{dz} + \frac{d\ln{K}}{dz} = \frac{dP}{dy} + \frac{d\ln{K}}{dy}\\
\!\!\!\! &\approx&\!\!\!\! -\frac{(y^3\!+\!6y\!+\!4)}{2(2\!-\!y)y(1\!+\!y)(2\!+\!y)}
 \!-\! \frac{(y^2\!+\!2y\!-\!2)(1\!+\!5\beta\!+\!3\beta^2\!+\!7\beta^3)}{16(1\!+\!y)(2\!+\!y)(4\!+\!y)
 (1\!+\!\beta)(1\!+\!\beta^2)},
\ea
and where I shall use
\be
z \equiv y-\s
\ee
for increased precision, since $\delta = 1-(a/M)^2$ is small, so that $z$, which vanishes for $\delta = 0$, is also small.

I could start with the solution above to the quadratic approximation for $\Delta\phi/\pi$, but as an alternative starting point, I shall use the quadratic approximation for $L(z)$, which I shall not derive in detail here but just write it out with its numerical coefficients, which depend on $K_m \equiv K(m_m) \equiv K([(\s-1)/2]^2)$ and $E_m \equiv E(m_m) \equiv E([(\s-1)/2]^2)$:
\be
L(z) \approx 0.010\,930\,048\,209 - 1.137\,078\,188\,23\, z - 0.610\,039\,842\,3\, z^2,
\ee
which is zero at what I shall take as the first iteration for $z$, at
\be
z_1 \approx 0.009\,563\,331\,457\,25.
\ee
This gives $y_1 = z_1 =\s \approx 1.423\,776\,893\,830\,35$, which is slightly smaller than the solution to the quadratic equation Eq.\ (\ref{quad}) in $\delta$.  It then leads to
\ba
L_1 \equiv L(z_1) &\approx& -0.000\,000\,785\,675\,4,\\
L_1' \equiv L'(z_1) &\approx& -1.148\,993\,606\,06.
\ea

Then Newton's method leads to
\be
z_2 = z_1 - L_1/L'_1 \approx 0.009\,562\,647\,662\,81,
\ee
which gives $y_2 \approx 1.423\,776\,620\,966\,07$ and $\Delta\phi/\pi \approx 1 - 5.1\times 10^{-13}$,
so just one iteration of Newton's method leads to a highly precise value for $z$ and $y = \s+z$ for getting $\Delta\phi = \pi$.  However, one can improve this estimate slightly by calculating (and using $L'_1$ as a sufficiently good approximation for $L'_2$ when I remain limited to using my 12-digit calculator)
\be
z_3 = z_2 - L_2/L'_2 \approx 0.009\,562\,647\,662\,63,
\ee
This leads to agreement to 12 significant digits (14 digits for $P$ and for $K$) for
\ba
1-\frac{1}{2}P &\approx& 0.010\,720\,888\,364\,8, \\
1 - \frac{\pi}{2K} &\approx& 0.010\,720\,888\,364\,8 ,
\ea
so $z_3$ seems to be about as close as I can get to the value for the photon boomerang, $\Delta\phi = \pi$ radians (180 degrees turning of the direction of the photon in one orbit when it gets back to the north polar axis), on my 12-digit HP 48X calculator.

One could also just start with the value for an extremal Kerr black hole, $z = y - \s = 0$ and use Newton's method with that initial value.  In that case it takes about three iterations to get precisions comparable to $z_2$ or $z_3$.

This value of $z = z_3$ gives the following values for other parameters of the photon boomerang black hole and constant-$r$ photon orbit that has a precisely reversed direction when it first returns to the axis from which it was emitted:
\ba
y \equiv \frac{r-M}{M} \equiv \s+z &\approx& 1.423\,776\,210\,035\,73, \\
\frac{r}{M} = 1+y = \s+1+z &\approx& 2.423\,776\,210\,035\,73, \\
\frac{r^2}{M^2} = (1+y)^2 = (\s+1+z)^2 &\approx& 5.874\,691\,116\,335\,2 , \\
\delta \equiv 1 - \frac{a^2}{M^2} = \frac{y(y^2-2)}{2+y} = \frac{(\s+z)(\sqrt{8}+z)z}{2+\s+z} 
&\approx& 0.011\,285\,617\,908\,8 , \\
\frac{a^2}{M^2} = \frac{(2-y)(1+y)^2}{2+y} = 1 -\delta &\approx& 0.988\,714\,382\,091\,2, \\
\frac{a}{M} = (1+y)\sqrt{\frac{2-y}{2+y}} = 1-\frac{\delta}{1+\sqrt{1-\delta}} 
&\approx& 0.994\,341\,179\,923\,26, \\
\frac{r^2+a^2}{M^2} = \frac{4(1+y)^2}{2+y} &\approx& 6.863\,405\,498\,426\,4 , \\
\frac{\Delta}{M^2} = \frac{2y(1+y)}{2+y} &\approx& 2.015\,853\,078\,354\,9, \\
\frac{\mathcal{K}}{M^2} = \frac{8(1+y)^3}{y(2+y)} &\approx& 23.367\,940\,622\,085, \\
P \equiv \frac{8Mar}{\Delta\sqrt{\mathcal{K}}} = (2+y)\sqrt{\frac{2(2-y)}{y(1+y)}} 
&\approx& 1.978\,558\,223\,270\,4, \\
m \equiv \frac{a^2}{\mathcal{K}} = \frac{y(2-y)}{8(1+y)} &\approx& 0.042\,310\,719\,550\,489\!, \\
1-m = \frac{(2+y)(4+y)}{8(1+y)} &\approx& 0.957\,689\,280\,449\,511\!, \\
\beta \equiv (1-m)^{1/4} &\approx& 0.989\,250\,222\,416\,6 , \\
\delta_2 \equiv \frac{1-\beta}{1+\beta}  &\approx& 0.005\,403\,934\,337\,8.
\ea

\section{Conclusions}

We have found that for a Kerr black hole with $a/M \approx 0.994\,341\,179\,923\,26$ (nearly maximally rotating), there is a constant-$r$ photon orbit at $r \approx 2.423\,776\,210\,035\,73\, GM/c^2$ that starts at the north polar axis and returns to it travelling in precisely the opposite direction, a photon boomerang.  Note that there is never a photon boomerang for null geodesics in any static metric, but for the stationary but nonstatic Kerr metric, null geodesic paths in space (projecting out the time part of the motion) are not invariant under reversing the spatial direction, so the twisting of the spacetime does allow a photon to come back travelling in the opposite direction to what it had when it was sent out.

Along the way to calculate the numerical values to high precision on just a pocket calculator, we have developed fairly compact closed-form formulas for the complete elliptic integrals $K(m)$ and $E(m)$ such that for the parameter values $m \leq [(\s-1)/2]^2$ that occur for constant-$r$ photon orbits through the polar axes, the relative error of $K(m)$ is less than $3\times 10^{-10}$ for $K_2(m)$ given in Eq.\ (\ref{K2}), $5\times 10^{-41}$ for $K_4(m)$ given in Eq.\ (\ref{K4}), $5\times 10^{-164}$ for $K_6(m)$ given in the unnumbered equation after Eq.\ (\ref{K4}), and $2\times 10^{-19}$ for the simplified $\bar{K}_4(m)$ given in  Eqs.\ (\ref{Kbar1})-(\ref{Kbar3}).  These formulas also have relative errors less than
$10^{-12}$ for $m$ less than $0.113$, $0.933$, $0.999\,999\,994$, and $0.276$ respectively, and relative errors less than 1\% for $m$ less than $0.978$, $0.999\,999\,999\,945$, $1-2.2\times 10^{-45}$, and $0.999$ respectively.  Therefore, the simplified $\bar{K}_4(m)$ given in Eqs.\ (\ref{Kbar1})-(\ref{Kbar3}) is more than adequate for high precision for the values of $m \leq [(\s-1)/2]^2 \approx 0.0429$ occurring in the photon boomerang problem, but the approximations $K_4(m)$ and $K_6(m)$ given in Eq.\ (\ref{K4}) and following are much better for significantly larger values of $m$.

As another comparison, we can take the value of the parameter very near 1, $m = \sin^2{\alpha} = \sin^2{(89^\circ 59' 59.5'')} \approx 1-(\pi/1\,296\,000)^2 \approx 0.999\,999\,999\,994\,123\,892\,365\,21$ (so, for example, the period of a simple pendulum of length $\ell$ would be $4\sqrt{\ell/g}K(m)$ with this $m$ when it is released from a position just one arcsecond from the top).  Then $\beta \equiv (1-m)^{1/4} \approx \sqrt{\pi/1\,296\,000} \approx 0.001\,556\,942\,004\,56$, and the approximations give $(2/\pi)K_0 = 1$ with 89\% relative error, $(2/\pi)K_2 \approx 3.987\,573\,492\,52$ with 56\% relative error, $(2/\pi)K_4 \approx 8.969\,080\,955\,34$ with 1.59\% relative error, $(2/\pi)K_6 \approx 9.114\,076\,326\,74$ with relative error $1.05\times 10^{-9}$, and $(2/\pi)K_8 \approx 9.114\,076\,336\,31$ with relative error $1.90\times 10^{-38}$.  Thus even the fairly short approximation $K_4(m)$ given by Eq.\ (\ref{K4}) gives the period of a pendulum with less than 1.59\% relative error over $647\,999/648\,000 \approx 99.999\,846\%$ of the range of initial positions for releasing the pendulum from rest.  For relative error less than $10^{-12}$, one needs $m$ to be less than about 0.87, which includes 67\% of the range of initial positions.  If one restricts the initial positions so that the bob is not higher than the pivot, which is half the range of initial positions, then $m \leq 1/2$, and the relative error of $(2/\pi)K_4$, which is $\approx 1.180\,340\,599\,02$ for $m = 1/2$, is less than $\approx 5.916\times 10^{-22}$, the approximate error at $m=1/2$ or $\alpha = \pi/4$ (initial pendulum bob position at angle $2\alpha = \pi/2 = 90^\circ$).

\section*{Acknowledgments}

This research was supported in part by the Natural Science and Engineering Council of Canada.

\end{document}